\def\bfig{\begin{figure}[h] \centerline{\hbox{}}\vfill}
\def\efig{\end{figure}\vfill\newpage}
\def\mycaption#1{\bigskip \parshape 1 1truecm 11.45truecm{#1}}

\def\ignore#1{}

\documentstyle[11pt,psfig]{article}

\begin{document}

\begin{center}

{\huge\sc Frontiers in High-Energy} 
\vskip0.2cm

{\huge\sc Astroparticle Physics}
\vskip0.5cm

{Karl Mannheim}
\vskip0.5cm

{\it Universit\"ats-Sternwarte}

{\it Geismarlandstra{\ss}e 11}

{\it D-37083 G\"ottingen}

{\it Germany}

\end{center}
\vskip0.3cm
\centerline{\bf Abstract}

\mycaption{\footnotesize{With the discovery of evidence for neutrino mass, a vivid gamma ray sky at
multi-TeV energies, and cosmic ray particles with unexpectedly high energies,
astroparticle physics currently runs through an era of rapid progress and moving
frontiers.   The non-vanishing neutrino mass establishes one
smooth component of 
dark matter which does not, however, supply a critical mass to the Universe.
Other dark
matter particles are likely to be very massive and should produce 
high-energy gamma rays, neutrinos, and protons in annihilations or decays.
The search for exotic relics with new gamma ray telescopes,
extensive air shower arrays, and underwater/-ice neutrino telescopes
is a fascinating challenge, but requires to understand the astrophysical
background radiations at high energies.  Among the 
high-energy sources in the Universe,
radio-loud active galactic nuclei seem to be the most powerful
accounting for at least a sizable fraction
of the extragalactic gamma ray flux.  They could also supply the bulk
of the observed cosmic rays at ultrahigh energies and produce interesting 
event rates in neutrino telescopes aiming at
the kubic kilometer scale such as
AMANDA and ANTARES.  
It is proposed that the extragalactic neutrino beam can be
used to search for tau lepton appearance thus allowing for a proof of the
neutrino oscillation hypothesis.
Furthermore, a new method
for probing the era of star formation at high redshifts using gamma rays is
presented which requires new-generation gamma ray telescopes operating in the
10-100~GeV regime such as MAGIC and GLAST.}}

\section{Introduction and practical definition of high-energy astroparticle
physics}

Central to modern astronomy is 
the dark matter problem and it is
commonly believed that its solution will trigger major advances in particle
physics and cosmology \cite{Turner99}.  
So far dark matter is only known through its
gravitational effects, but the understanding of the nature and origin
of dark matter requires to obtain more direct information about its mass and
interactions.  Cosmology and particle physics qualify weakly interacting massive
particles (WIMPs) with masses between 100~GeV and a few TeV as likely
candidates.  The WIMPs violently
annihilate with their anti-particles in rare collisions or they could be
unstable.  These modes lead to secondary gamma rays and neutrinos
which can be detected on Earth \cite{Gaisser95}.  The suspected large WIMP
mass then corresponds to gamma ray and neutrino energies in excess of 100~GeV.
In addition to the dark matter particles, there may be other relics from the
early Universe, such as quintessence, vacuum energy, or topological defects.  
Quintessence is a slowly rolling scalar field connected with massive
bosons, and this could also lead high-energy phenomena, although
no worked-out model exists to my knowledge.  The recent discovery of 
accelerating expansion using SNIa as a tracer of cosmic geometry seems to
make a strong case for quintessence \cite{Perl98,Schmidt98}.  
Similarly, if the scalar field has settled
to some false (meta-stable) vacuum, the energy density of this vacuum could
drive the deceleration parameter away from $\Omega/2$.  Topological defects
preserve false vacuum  in pointlike (monopoles), one-dimensional (strings),
two-dimensional (domain walls), or higher dimensional space-time structures.
They are topologically stable, but have a variety of ways to communicate
to our world in addition to their gravitation.  E.g., they
can dissipate into GUT-scale bosons ($\sim 10^{16}$~GeV)
which are unstable
themselves and fragment into jets consisting of gamma rays, neutrinos, and
protons at ultra-high energies \cite{Sigl94}.  
After their propagation through intergalactic
space, electromagnetic cascading and secondary particle production shift most
energy injected by these exotic processes to much lower energies where the
energy release competes with that due to ordinary astrophysical sources.\\

{\em In order to identify new physics phenomena, it is therefore
of crucial importance to 
obtain a complete inventory of the astrophysical high-energy sources which
act as a background for these searches.  This defines the high-energy
astroparticle physics program from a practical point of view and 
follows  the logic inherent to the
general astronomical exploration of the sky to cover the entire range
of wavelengths with comparable sensitivity.}\\

Among the non-thermal sources in the Universe, radio-loud active galactic nuclei
(AGN) seem to be the most important energetically.  There are other interesting
sources, such as gamma ray bursts (GRB) and clusters of galaxies, but their
non-thermal energy release does not come close that of AGN.  There are some
intriguing complications arising through calorimetric effects, since the
intracluster medium in clusters of galaxies surrounding AGN confines escaping
relativistic particles for some time and thereby gives rise to a secondary
luminosity tied to the energy release of the AGN and the cooling time scale of
the intracluster medium.  The radio-loud AGN come in various disguises,
depending on the orientation of their radio jet axes and the properties of the
circum-nuclear matter in their host galaxies.  The most extreme versions are
radio galaxies with the radio jet axes almost in the plane of the sky and the
blazars with the radio jets pointing close to the line of sight to the observer
which leads to a dramatic flux increase owing to special relativistic effects
(the so-called Doppler boosting).  In Sect.2 it is
argued that radio-loud AGN can be expected to
produce the entire extragalactic gamma ray
background from an energetical point of view.  Owing to beaming statistics, the number of unresolved sources
responsible for most of this background should not be too large.  Indeed the
flux from the $\sim 50$ resolved sources in the flux-limited EGRET sample
already equals a sizable fraction (of the order of 15\%) of the
extragalactic background flux.  
With the next-generation gamma ray telescopes MAGIC \cite{MAGIC}
(giant 17m air-Cerenkov telescope) and GLAST \cite{GLAST}
(space-borne silicon strip
detector) it will be possible to probe deeper into the astrophysical source
population producing the extragalactic background below 100~GeV thereby
narrowing the range of opportunity for particle physics models of exotic
processes producing gamma rays.  In fact, the gamma ray background below 100~GeV
provides a good measure of the entire electromagnetic energy release during the
history of the Universe, since gamma rays from
remote sources cascade down to the energy range below
100~GeV which is shown in Sect.3.  It is pointed out in Sect.4 that a MAGIC
observing campaign for high-redshift gamma ray sources can be used to probe the
era of star formation and the evolution of the optical-ultraviolet metagalactic
radiation field back to redshifts of $\sim 5$.\\

The recent discovery of multi-TeV emission from Mrk~421 and Mrk~501 
\cite{Aharon97},
the measurement of their spectra using the HEGRA air-Cerenkov imaging
telescopes \cite{Konol99}, and the improved
measurements of the extragalactic infrared background \cite{Dwek99},
make a strong point in favor of accelerated protons
in extragalactic radio sources which is shown in Sect.5.
The following Sect.6 discusses the
immediate implications that the radio-loud AGN could well produce
the observed cosmic rays at highest energies and high-energy muon neutrinos.
Finally, in Sect.7 it is pointed out that the extragalactic muon neutrino
beam is likely mixed with tau neutrinos \cite{SuperK98}
which leads to very interesting
experimental signatures, such as the disappearance of the Earth 
shadowing effect at ultra-high energies and the appearance of tau
leptons in underwater/-ice detectors.

\section{Origins of extragalactic background radiation}

Inspection of Fig.~1 shows an interesting pattern
in the present-day energy
density of the diffuse isotropic background radiation 
consisting of a sequence of bumps each with a strength that is
decreasing with photon energy.
The microwave bump is recognized as the signature of the big bang
at the time of decoupling with its energy density given by
the Stefan-Boltzmann law $u_{\rm 3K}=a T^4$.
The bump in the far-infrared is due to star formation in early galaxies, 
since part of the stellar light, which is visible
as the bump at visible wavelengths, is reprocessed by
dust obscuring the star-forming regions.  The energy density of the two bumps
can be inferred from the present-day heavy element abundances.
Heavy elements have a mass fraction 
$Z=0.03$ of the total mass density $\rho_*$
and were produced in early bursts of star formation at redshift $z_{\rm f}$
by nucleosynthesis with radiative efficiency 
$\epsilon=0.007$ yielding
\begin{equation}
u_{\rm ns}\sim {\rho_* Z \epsilon c^2\over  1+z_{\rm f}}\ .
\end{equation}
Inserting plausible parameter values one obtains
\begin{equation}
u_{\rm ns}\sim
6\times 10^{-3}\left(\Omega_*h^2\over  0.01\right)\left(1+z_{\rm f}\over
3\right)^{-1}\ {\rm eV~cm^{-3}}
\end{equation}
for the sum of the far-infrared and optical bumps.
Probably all galaxies (except dwarfs) contain
supermassive black holes in their centers which are actively
accreting over a fraction of $t_{\rm agn}/ t_*\sim 10^{-2}$
of their lifetime implying that the electromagnetic radiation
released by the accreting black holes amounts to
\begin{equation}
u_{\rm accr}\sim {\epsilon_{\rm accr}M_{\rm bh}\over  Z\epsilon M_*}
{t_{\rm agn}\over  t_*}u_{\rm ns}\sim 1.4\times 10^{-4}\ \rm eV~cm^{-3}
\end{equation}
adopting the accretion efficiency $\epsilon_{\rm accr}=0.1$
and the black hole mass fraction $M_{\rm bh}/M_*=0.005$
\cite{Rees98}.  Most of the accretion power emerges in
the ultraviolet where the diffuse background is unobservable
owing to photoelectric absorption by the neutral component of the
interstellar medium.  However, a fraction of $u_{\rm x}/u_{\rm bh}\sim
20\%$ taken from the average quasar spectral energy distribution
\cite{Sanders89} shows up in hard X-rays due to coronal emission 
from the accretion disk to produce the diffuse isotropic X-ray background
bump with 
\begin{equation}
u_{\rm x}\sim 2.8\times 10^{-5}~\rm eV~cm^{-3}
\end{equation}
\cite{Gruber92}.
\begin{figure}
\centerline{\psfig{figure=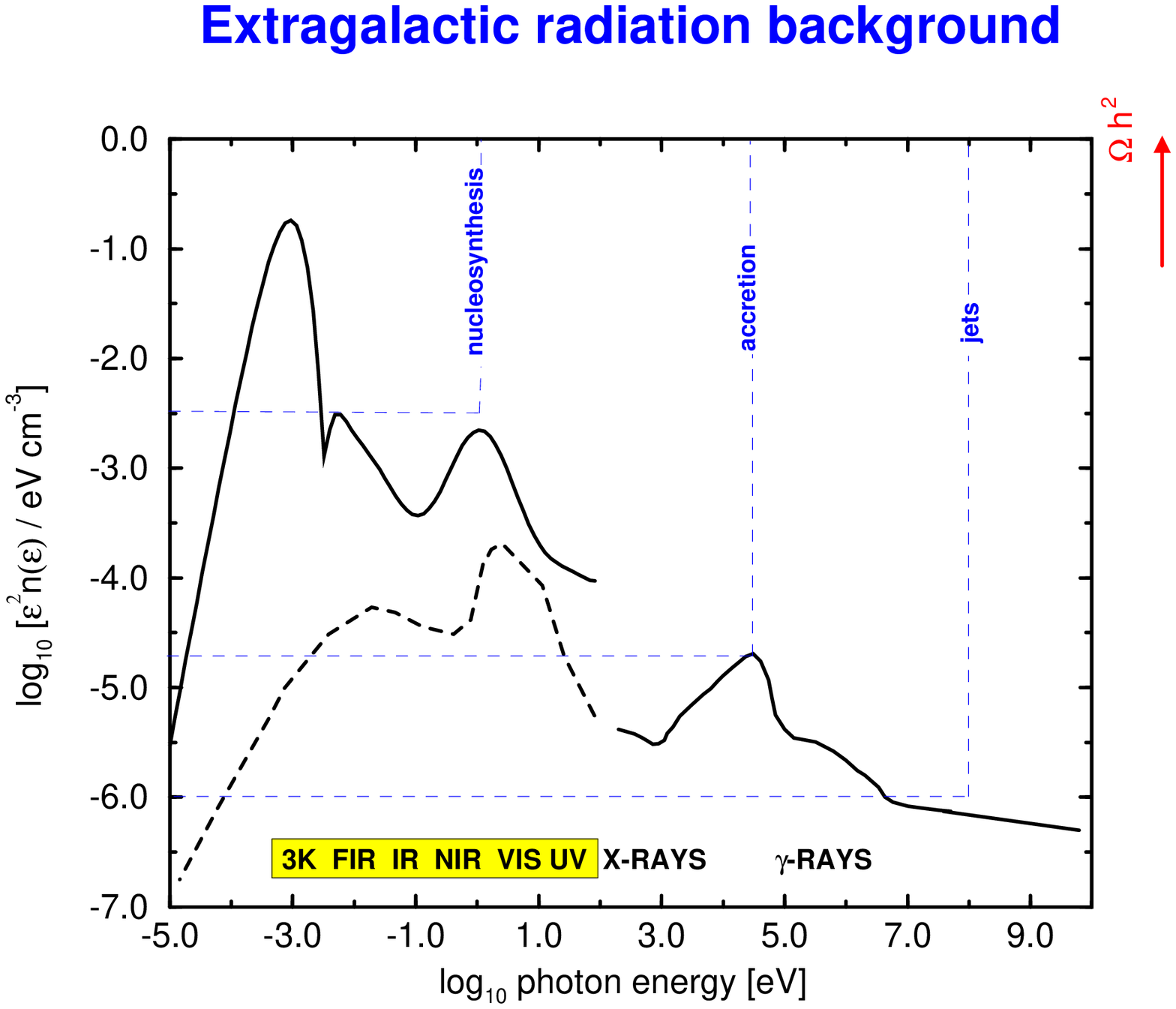,width=10cm,height=8cm}}
\mycaption{{\bf Fig.1:} Sketch of the present-day energy density
of the extragalactic radiation background from radio waves to
gamma rays. The dashed line shows the expected AGN contribution
to the low-energy diffuse background from the average quasar spectral
energy distribution.}
\label{diffuse_entire}
\end{figure}

Jets with non-thermal $\gamma$-ray emission show up 
only in the radio-loud fraction $\xi_{\rm rl}\sim 20\%$ of
all AGN and their kinetic power roughly equals
the accretion power \cite{RS91}.  Hence
one obtains for the background energy density due to extragalactic jets
\begin{equation}
u_{\rm j}=\left(\xi_{\rm rl}\over 0.2\right)u_{\rm accr}\sim 
\left(\xi_{\rm rl}\over 0.2\right)2.8
\times 10^{-5}~\rm eV~cm^{-3}
\end{equation}
If unresolved extragalactic jets are
responsible for the 
diffuse gamma ray background, this requires
a particle acceleration efficiency given by
\begin{equation}
\xi_{\rm acc}={u_{\rm acc}\over
u_{\rm j}}= {u_\gamma \over \xi_{\rm rad} u_{\rm j}}
\end{equation}
where $u_{\rm j}$ denotes the total (kinetic + magnetic + randomized
relativistic particle) energy density in extragalactic jets. 
Inserting  the energy density
of the observed extragalactic gamma ray background\footnote{
Note that the flux in the gamma ray background observed by CGRO is close
to the bolometric gamma ray flux of the Universe,  since pair attenuation and cascading
must lead to a steepening of the background spectrum above $20-40$~GeV
\cite{MP96,Salamon98}.}
one obtains a limit
for the acceleration efficiency
\begin{equation}
\xi_{\rm acc}\ge 0.18\xi_{\rm rad}^{-1}
\left(\xi_{\rm rl}\over 0.2\right)^{-1}
\end{equation}
which is of the same order of magnitude as the 13\% efficiency required for
supernova remnants to produce the Galactic cosmic
rays.  Accelerated protons
achieve this high radiative efficiency, if 
they reach energies of up to $10^{8}$~TeV.
In the next section it is shown that protons at such high energies
cannot go unnoticed, they produce interesting gamma ray spectra
owing to the photo-production of secondaries.  Some of the protons
turn into neutrons due to $\pi^+$ production and can leave the
jets without adiabatic losses.  These particles would have just
the right flux to produce the observed extragalactic cosmic rays
dominating the local spectrum above $10^{18.5}$~eV, as will be shown
in Sect.6.

\section{Cascading and gamma ray calorimetry}

Gamma rays of energy $E$
can interact with low-energy
photons of energy $\epsilon$ 
from the diffuse isotropic background over cosmological distance
scales $l$ producing electron-positron pairs
$\gamma+\gamma\rightarrow e^++e^-$, if their energy exceeds the
threshold energy
\begin{equation} 
\epsilon_{\rm th}={2 (m_{\rm e}c^2)^2\over
(1-\mu)(1+z)^2E}\sim 1\left(1+z\over 4\right)^{-2}\left(E\over 30~{\rm GeV}
\right)^{-1}~{\rm eV}\ \ (\mu=0) 
\end{equation} 
where $\mu$ denotes the cosine of the scattering angle
\cite{gould66}.
The $\gamma$-ray attenuation $e^{-\tau}$ 
due to pair production becomes important
if the mean free path
$\lambda$ becomes smaller than $l$, i.e.
if the optical depth across the line of sight through a sizable fraction
of the Hubble radius obeys $\tau=l/\lambda\ge 1$.
For the computation of $\tau$
one first needs to know the pair production cross section 
\begin{equation}
\label{sigma}
\parbox[b]{10cm}{}
\sigma_{\gamma\gamma}={3\sigma_{\rm T}\over 16}
(1-\beta^2)\left[2\beta(\beta^2-2)+(3-\beta^4)\ln\left(1+\beta\over
1-\beta\right)\right]
\end{equation}
where $\beta=\sqrt{1-1/\gamma^2}$ with $\gamma^2=\epsilon/\epsilon_{\rm th}$,
and where $\sigma_{\rm T}$ denotes the Thomson cross section
\cite{jauch76}.
Then one needs 
the geodesic radial displacement
function 
$dl/dz={c\over H_\circ}[(1+z)\bar{E}(z)]^{-1}$  
to compute the line integral from
$z=0$ to some $z=z_\circ$. For a cosmological model with $\Omega=1$
and $\Lambda=0$ the function $\bar{E}(z)$ simplifies to $(1+z)^{3/2}$.
Hence one obtains the optical depth
\begin{eqnarray}
\label{tau}
\tau_{\gamma\gamma}(E,z_\circ)=
\int_0^{z_\circ}dz{dl\over dz}\int_{-1}^{+1}d\mu{1-\mu\over 2}
\int_{\epsilon_{\rm th}}^\infty
d\epsilon n_{\rm b}(\epsilon)(1+z)^3\sigma_{\gamma\gamma}(E,\epsilon,\mu,z)
\nonumber\\
={c\over H_\circ}\int_0^{z_\circ}dz
(1+z)^{1/2}\int_{0}^{2}dx{x\over 2}\int_{\epsilon_{\rm th}}^\infty
d\epsilon n_{\rm b}(\epsilon)\sigma_{\gamma\gamma}(E,\epsilon,x-1,z)
\end{eqnarray}
adopting a non-evolving present-day background density $n_{\rm b}$,
i.e. 
$n_{\rm b}'(z,\epsilon')d\epsilon'
=(1+z)^3n_{\rm b}(\epsilon)
d\epsilon$ where the dash indicates comoving-frame quantities. 
The simplifying assumption  that the photon density transforms 
geometrically 
corresponds to the situation in which an initial short burst of star
formation at $z_{\rm f}>z_\circ$ produced most of the diffuse 
infrared-to-ultraviolet background
radiation.  This simple assumption is replaced by a more realistic
one in Sect.4. 
Fig.2 shows the spectrum of the low-energy diffuse background
used to solve Eq.(10) numerically.

\begin{figure}
\centerline{\psfig{figure=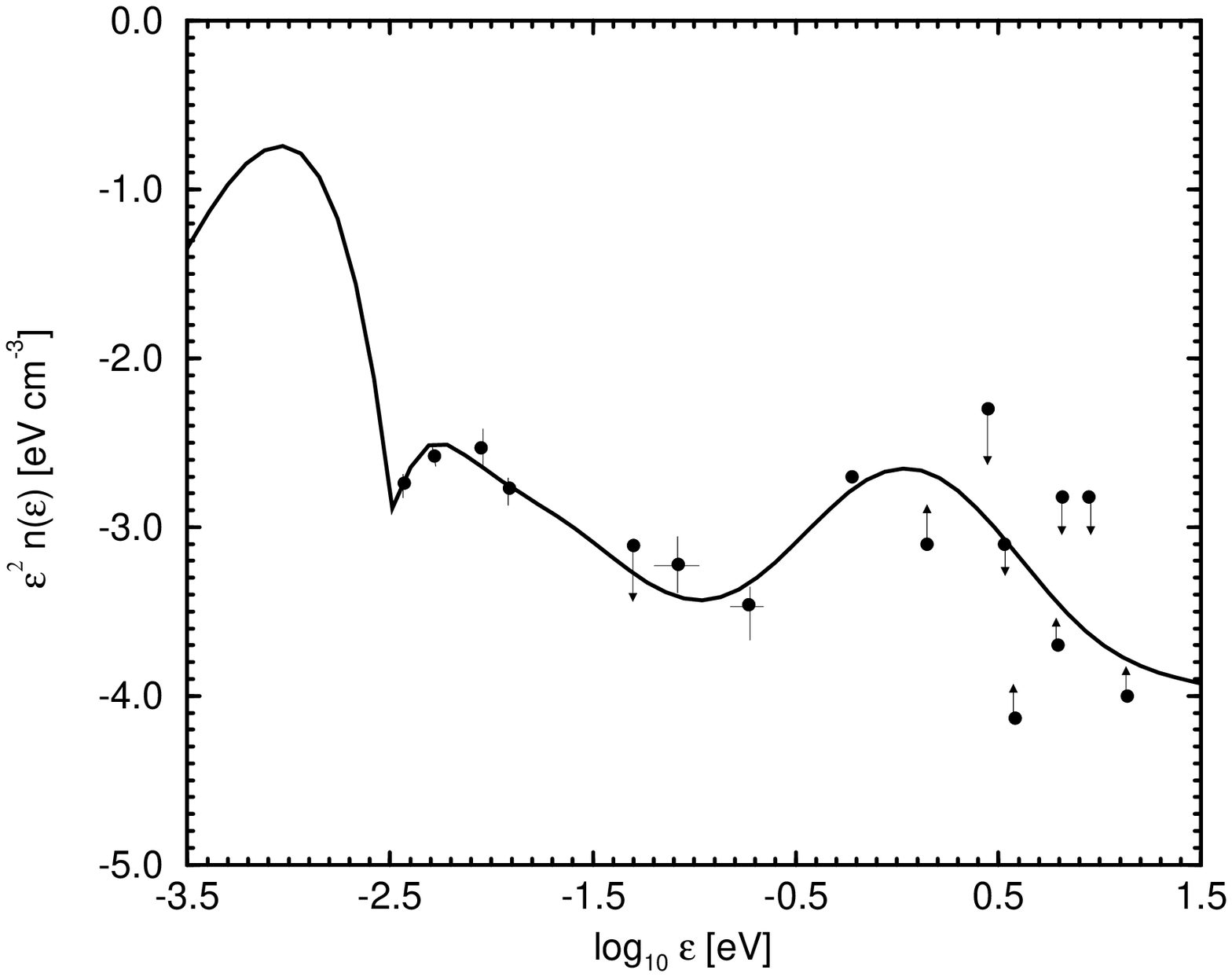,width=10cm,height=8cm}}
\mycaption{{\bf Fig.2:} 
The diffuse isotropic microwave-to-ultraviolet background. Solid
curve: 10th order polynomial interpolation of
observational data (\cite{fixsen98,mannheim98,franceschini98}, 
and references in \cite{MP96}).}
\end{figure}

\begin{figure}
\centerline{\psfig{figure=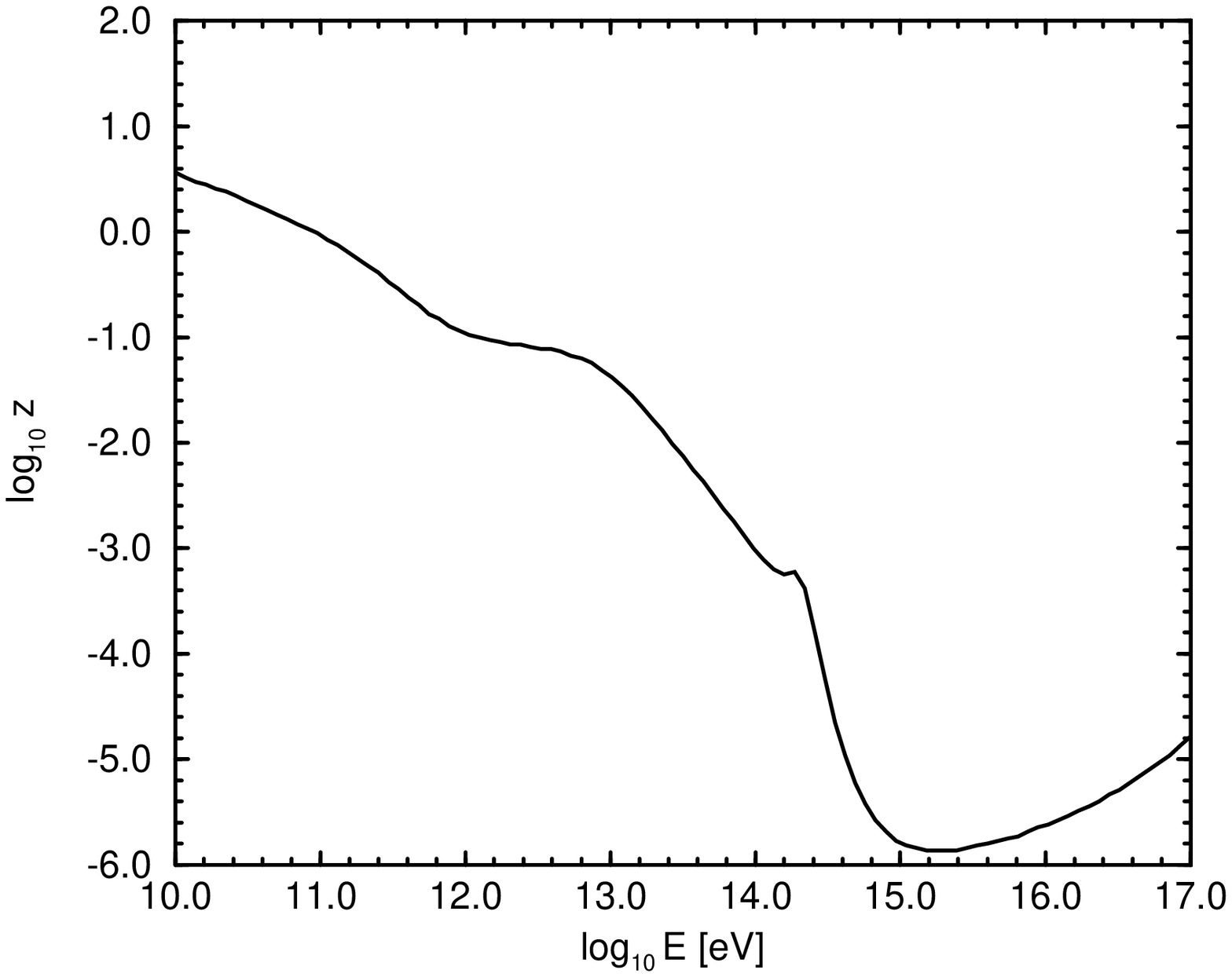,width=10cm,height=8cm}}
\mycaption{{\bf Fig.3:} 
The $\gamma$-ray horizon $\tau(E,z)=1$ for the low-energy
background spectrum shown in Fig.2.  Cosmological parameters are
$h=0.6$, $\Omega=1$, and
$\Omega_\Lambda=0$. For a general discussion of pair attenuation,
see reference \cite{biller96}.}
\end{figure}
Figure 3 shows the resulting $\tau(E,z)=1$ (omitting the subscript hereafter)
curve for the microwave-to-ultraviolet diffuse background spectrum shown in
Fig.2.  It is obvious that $\gamma$-rays above $\sim 10-50$~GeV cannot reach
us from beyond redshifts of $z=z_{\rm f}=2-4$.  Higher energy $\gamma$-rays
can reach us only from sources at lower redshifts (e.g.  $\gamma$-rays with
energies up to 10~TeV have been observed from Mrk~501 at $z=0.033$ in accord
with Fig.3 \cite{mannheim98}).

\vskip0.5cm
{\bf Corollary I:}  {\em If the 
extragalactic gamma ray background
originates from unresolved sources
distributed in redshift similar to galaxies, its spectrum must steepen above
$\sim 30$~GeV due to $\gamma$-ray pair attenuation.}
\vskip0.5cm

Here is has been tacitly assumed that the $\gamma$-rays which have turned into
electron-positron pairs do not show up again.  This is, in fact, not quite true,
since the pairs are subject to inverse-Compton scattering off the microwave
background thereby replenishing $\gamma$-rays.  The 2.7~K background is more
important as a target than the shorter wavelength background, since there is no
threshold condition for Thomson scattering 
contrary to pair production and since
2.7~K photons greatly outnumber the latter.  The inverse-Compton scattered
microwave photons turn into $\gamma$-rays of energy
\begin{equation}
E_{\rm ic}\sim 10\left(1+z\over
4\right) \left(E\over 30~{\rm GeV}\right)^2 \ {\rm MeV}
\end{equation}
conserving the energy of the absorbed $\gamma$-ray
which corresponds to a constant $E^2dN/dE$,
i.e. the expected slope of the differential spectrum
is about -2 (-2.1 observed).
A small amount of energy is lost to lower frequency synchrotron
emission, if magnetic fields are present in the interagalactic medium.

\vskip0.5cm
{\bf Corollary II:}  {\em Energy conservation in the reprocessing of
$\gamma$-rays from higher to lower energies by pair production and subsequent
inverse-Compton scattering produces an approximate $dN/dE\propto E^{-2}$ power
law extragalactic gamma ray
background between $\sim 10$~MeV and $\sim 30$~GeV.}
\vskip0.5cm

\section{Evolution of the metagalactic optical-ultraviolet radiation field}

\subsection{A simple model based on the observed "effective" star formation
rate}

Consider an effective\footnote{The term ``effective'' means that only the
star formation rate inferred from photons which have made it through
possible obscuring dust clouds are of relevance for the build-up of
a metagalactic radiation field.}
cosmic star formation history $\dot\rho_*(z)$ denoting the production
rate per unit volume of mass which has formed to stars at a redshift of z.  Such
star formation histories have been inferred from galaxy counts in the Hubble Deep
Field \cite{Madau}.
Since the present-day infrared background is strong enough to aborb
gamma rays in the TeV range in the local Universe far from the peak of the star
formation history, its evolution in the past is rather irrelevant in this
context.  However, the present-day optical radiation background scaled back to
the peak of the star formation history at a redshift of $z_{\rm b}\sim 1.5$
implies gamma ray attenuation in the 20~GeV regime from sources at this
redshift.  Over this distance scale the evolution of the background becomes
important, since it is gradually produced by the forming stars.\\

For the gamma ray attenuation only the evolution of the number density
$n(\epsilon,z)d\epsilon$ of the background photons is important which relates to
the photon production rate $\dot n\propto \dot \rho/\epsilon$ in the following
way:

\begin{equation}
n(\epsilon,z)d\epsilon=\int_z^{z_{\rm f}}dz'\dot n(\epsilon',z')
{dt\over dz'}\left(1+z'\over 1+z\right)^{-3}d\epsilon'
\end{equation}

The evolving background must be normalized to yield the observed
present-day radiation background

\begin{equation}
n(\epsilon,0)d\epsilon=\int_0^{z_{\rm f}} dz' 
\dot n(\epsilon',z')
{dt\over dz'}\left(1+z'\right)^{-3}d\epsilon'
\end{equation}

As a simple example consider a burst of star formation at a high
redshift $\dot n\propto \delta(z-z_{\rm f})$.  Inserting this in
Eq.(12) and combining with Eq.(13) we obtain

\begin{equation}
n(\epsilon,0)d\epsilon=(1+z)^{-3}n(\epsilon',z)d\epsilon' \ \ \ 
(z\le z_{\rm i})
\end{equation}

which 
represents a constant co-moving density background density where the
$(1+z)^3$ term reflects the geometric scaling of the cosmic volume.  To obtain a
realistic parametrization of $\dot n(z,\epsilon')d\epsilon'$ we approximate the
Madau curve \cite{Madau} as a broken power law

\begin{equation}
\dot n(z,\epsilon')(1+z')^{-3}\propto \left(\dot\rho_*(z)\over
\epsilon \right)\propto (1+z)^{\alpha-1}
\end{equation}

with $\alpha=\alpha_{\rm M}=3.8$ for $0\le z\le 1.5=z_{\rm b}$ and
$\alpha=\beta_{\rm M}=-4.0$ for $z_{\rm b}=1.5\le z\le 10=z_{\rm f}$.
We also investigate a star formation rate which exhibits a plateau
beyond $z_{\rm b}$.

Equation (1) enters the formula for the gamma ray optical depth:
\begin{equation}
\tau(E_\circ, z)=\int_0^z dz {dl\over dz}\int_{-1}^{+1}d\mu (1-\mu)
\int_{\rm \epsilon_{\rm th}}^\infty d\epsilon n(\epsilon,z)
\sigma(E,\epsilon,\mu)
\end{equation}

Note that cosmology enters through $dl/dz$ (which depends on
$\Omega,\Omega_\Lambda,$ and $H_\circ$), not through $n(\epsilon,z)$
for a given parametrization of $\dot\rho_*(z)$.  
However, the parametrizations $\dot\rho_*(z)$ must also
satisfy observational constraints such as number counts and
the present-day diffuse background
which themselves depend on cosmology.
Turning this around it means that one must find the cosmology parameters 
for which a measured gamma ray horizon
(i.e., the curve $\tau(E_\circ,z)=1$) and the star formation history
data come into mutual consistency.

The gamma ray horizon from Eq.(16) is shown in Fig.4 using the
low-energy background spectrum template shown in Fig.2.  The
template was used to normalize the evolving background such
that is identical to the template at $z=0$ and scales to
higher redshifts acording to Eq.(12).
The fact that there is a redshift with a maximum star formation is
very important.  If power law evolution of the background
emissivity were to continue all the way in the past, one could
easily infer power law solutions for the scaling of
$n(\epsilon,z)$ which are more shallow than $(1+z)^3$
as in ref. \cite {MP96},
but such solutions become unrealistic beyond $z_{\rm b}$.

\begin{figure*}
\centerline{\psfig{figure=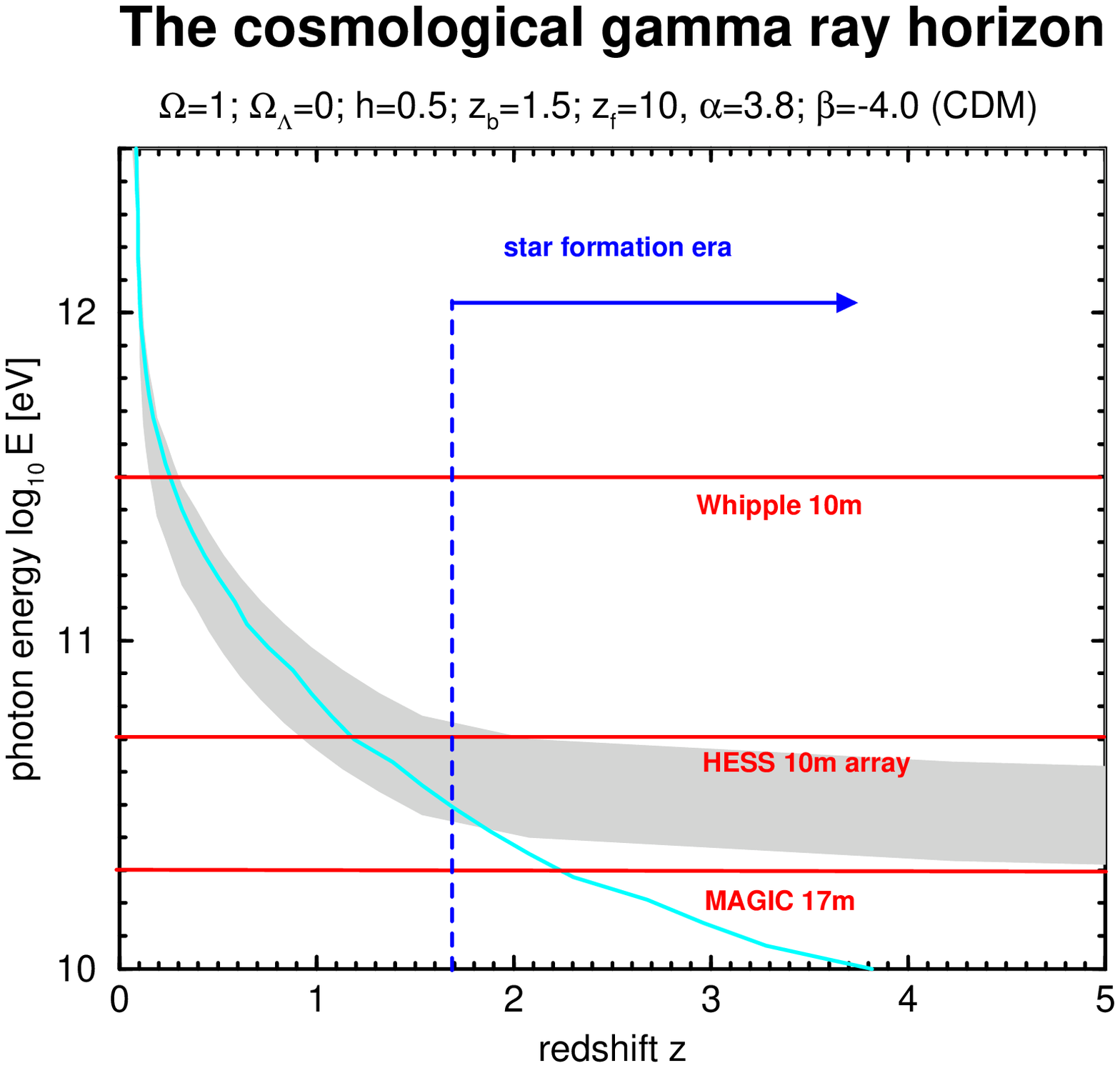,width=12cm,height=10cm}}
\mycaption{{\bf Fig.4:} 
Gamma ray horizon due to interactions with an evolving metagalactic
radiation field as computed from the effective star formation
rate.  Sources below the horizon curve suffer no significant 
pair-attenuation along the line of sight.
The grey band indicates the uncertainty of the horizon 
as estimated from the range
of interpolations allowed between observational upper and lower bounds of
the flux of the present-day optical-UV diffuse
background.
Note that the metagalactic radiation field before the maximum
of the cosmic star formation rate (indicated by the dashed line)
is too weak to significantly 
attenuate gamma rays below 20-40~GeV resulting in a near-constant
optical depth.  The light solid line shows the effect of a
star formation rate with an extended plateau
which causes the optical depth to continue
to grow with redshift beyond $z=1.5$. The horizontal lines indicate
the effective threshold energies for various air-Cerenkov telescopes.
It is emphasized that triggering below 20-40~GeV, which can
be achieved by the MAGIC telescope, is crucial for probing
the star formation era.  Such an investigation is complementary to 
studies of galaxies at high redshifts, since it is additionally sensitive to 
diffuse sources of optical-UV photons such as would be arising from 
exotic particle decays (one possible scenario for the reionization epoch).}
\label{horizon}
\end{figure*}

\subsection{Extragalactic gamma ray background}

The origin of the observed diffuse isotropic gamma ray background
is unknown.  The spectral shape and flux density suggest that 
unresolved faint radio loud AGN are responsible for this background,
similar to the situation in the X-ray band where deep observations
have revealed that faint AGN are responsible for more than 90\% 
of the background emission. 
The EGRET-type radio-loud AGN seem contribute not more than $\sim 25\%$ 
\cite{Chiang}
to the extragalactic gamma ray background.  The uncertainties 
about the faint end of the gamma ray luminosity function in the EGRET band
allow for a larger contribution from the general class of radio-loud AGN.  
According to beaming statistics, the flux-limited EGRET sample of AGN is
dominated by highly beamed sources with a rather flat luminosity function.
A much fainter, less beamed population with a steeper luminosity function is
likely to fill in the remaining 75\%, at least this seems very plausible considering the
energetics of radio jets as was shown in Sect.2.  I strongly expect that the
flat-spectrum/steep-spectrum classes are mirrored in different populations
of gamma ray sources.  The nearest steep-spectrum radio sources would
have been detected by EGRET even if they were faint (with their gamma ray
luminosity roughly a factor of $\sim$ 50 larger than the 5 GHz luminosity).
However, with a lower compactness for instrinsic gamma ray absorption
$\propto L/R$ the steep-spectrum radio sources could emit most of their
gamma ray power above the EGRET range.  It is up to new air-Cerenkov
telescopes with threshold energies above 10~GeV and GLAST to probe this proposal.\\

If the extragalactic jets are indeed responsible for most of the gamma ray background,
it is straightforward to investigate the effect of pair attenuation on the spectrum of
the background.  The precise shape of the spectra of the individual sources
at high redshifts is rather unimportant owing to the effects of cascading discussed in Sect.3.
Adopting a power law gamma ray spectrum per source with the average
slope of the resolved EGRET sources 
\begin{equation}
{dN\over dE}=A \left(E\over E_1\right)^{-2.1}
\end{equation}
extending from $E_1=10$~MeV to $E_2=1$~TeV
 and taking into account the luminosity density
evolution $\Psi(z)$ of AGN, we obtain a good approximation of
the present-day background energy density 
\begin{equation}
u(E)={4\pi\over c}E I_E
\end{equation}
from the equation
\begin{equation}
u(E)\propto \int_0^{z_{\rm f}} dz {dt\over dz}\Psi(z)(1+z)^{-4}
B\left(E(1+z)\over E_1\right)^{-0.1}e^{-{E(1+z)\over E_2}}
C[E,z]
\end{equation}
where $dt/dz$ is given by
\begin{equation}
dt/dz={1\over (1+z)\sqrt{\Omega(1+z)^3+(1-\Omega-\Omega_\Lambda)
(1+z)^2+\Omega_\Lambda}}
\end{equation}
and the function $C[E,z]$ for the effect of pair attenuation
can be approximated as 
\begin{equation}
C[E,z]=e^{-{E\over E_{\rm t}(z)}}
\end{equation}
with $E_{\rm t}(z)$ denoting the solution of the
equation $\tau(E,z)=1$ (the gamma ray horizon).
The result is shown in Fig.~5.  It will be possible
with GLAST to find whether the diffuse gamma ray background
indeed turns over in this shallow fashion or continues as
a power law into the 100~GeV domain.  In the latter case,
the gamma ray background would have to be due to some
local source population \cite{Dar99}.

\begin{figure*}
\centerline{\psfig{figure=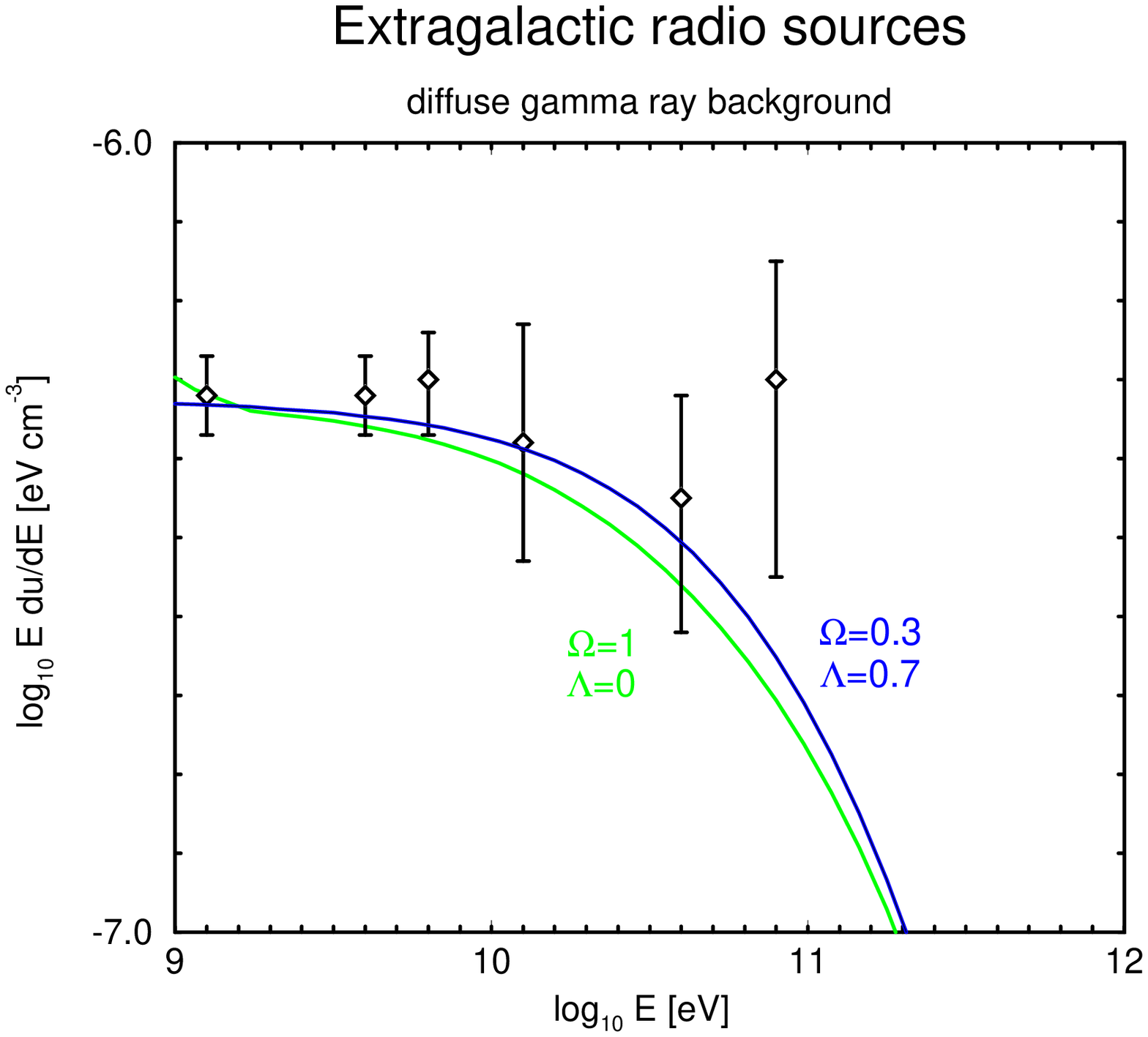,width=10cm,height=8cm}}
\mycaption{{\bf Fig.5:} 
The gamma ray horizon 
enforces a shallow turnover beyond $\sim 30$~GeV in the spectrum
of the extragalactic gamma ray background which is only weakly 
dependent on cosmology (results from further investigations by T. Kneiske
will soon be reported elsewhere.). }
\label{diffuse_gamma}
\end{figure*}

\section{Comparison of proton blazar predictions with observed multi-TeV
spectra}

In the previous sections arguments based on energetics have been used
to favor extragalactic jets as the sources of the gamma ray background.
This requires that the jets radiate a sizable fraction of their kinetic energy
in gamma rays when integrated over their lifetimes.  Since the cooling of
relativistic particles increases with their energy, a high radiative efficiency
is equivalent with high energies.  For electrons, Lorentz
factors required for a high radiative efficiency are at least $\gamma_e\sim 10^3$,
and for protons $\gamma_p\sim 10^9/(1+500 u_\gamma/u_B)\gamma_e
\sim 10^{10}$.
The Lorentz factor for protons may seem outrageously high, but in a statistical
acceleration process such as Fermi acceleration
with the balance between energy gains and losses
determining the maximum energy, such high energies are an inevitable consequence
of the acceleration theory.  Moreover, particles with energies in excess of
$10^{19}$~eV are observed in the local spectrum of cosmic rays.  Their
energy is too high for the gyrating particles
to be isotropized in the Galactic disk, so that an extragalactic
origin is very likely.  The energy requirements can be converted to an energy
supply rate for extragalactic sources, and this requires sources as strong as 
radio galaxies.  Thus, if it is not the radio sources themselves, some other
source must be able to produce a relativistic proton distribution with an 
enormous energy flux reaching these ultrahigh-energies.  It has been suggested
that GRBs may do that, but there are strong arguments against that proposal
presented in Sect.6.  Here we concentrate on the consequences of 
accelerating baryons in radio jets to these extremely high energies where
radiative losses become important.  A few years ago, a model coined the
proton blazar model has been presented in which
accelerated protons have been assumed in addition
to the accelerated electrons \cite{MKB91}.  This model made definite
predictions about the multi-TeV spectra of nearby blazars \cite{man96}
which can now be compared
with observations of Mrk~501 and Mrk~421 obtained with the HEGRA
air-Cerenkov telescopes.  In fact, this was the only model prior to the
observations which made {\em any} quantitative prediction of multi-TeV emission
from these sources.\\

An important, although not necessary,
 assumption of the original model is that the magnetic field pressure
energy density is in equipartition
with that in relativistic particles implying that synchrotron cooling
dominates over Compton cooling (since the photon energy density
remains below that in particles).  
This affects accelerated electrons as well as secondary electrons at 
ultrahigh-energies.
Evaluating
a simple conical jet geometry shows that typical blazars
are optically thin to gamma rays
up to the TeV range.  Unsaturated synchrotron
cascades initiated by accelerated protons interacting
with the synchrotron photons from the accelerated electrons
are computed
as the stationary solution of a coupled set of kinetic equations
which is then Doppler boosted to an appropriate observer's frame.
A series solution is found employing Banach's fixed point theorem
which can be physically interpreted as a series of superimposed 
cascade generations.  The cascade generation of gamma rays 
emerging in the TeV range on the optically thin side has
a spectral index $s\sim 1.7$ (differential spectrum $I_\circ$)  
steepening by $\alpha=0.5-0.7$ above TeV.  The index $\alpha$ is
the energy index of the
optical synchrotron photons which act as a target for both
gamma rays and protons. The reason for the break is
the onset of intrinsic pair attenuation
characterized by the escape probability $I_\gamma=P_{\rm esc}I_\circ$
for
a homogeneously mixed absorber
and emitter 
\begin{equation}
P_{\rm esc}={1-\exp[-\tau(E)]\over  \tau(E)}
\rightarrow {1\over \tau(E)}
\propto E^{-\alpha}\ \ \ \ \ {\rm for\ \ } \tau\gg 1
\end{equation}
The shape of the multi-TeV spectrum is therefore not sensitive to changes
in the maximum energy and can remain constant under large-amplitude
changes of the flux associated with changes in the maximum energy.\\

\begin{figure*}
\centerline{\psfig{figure=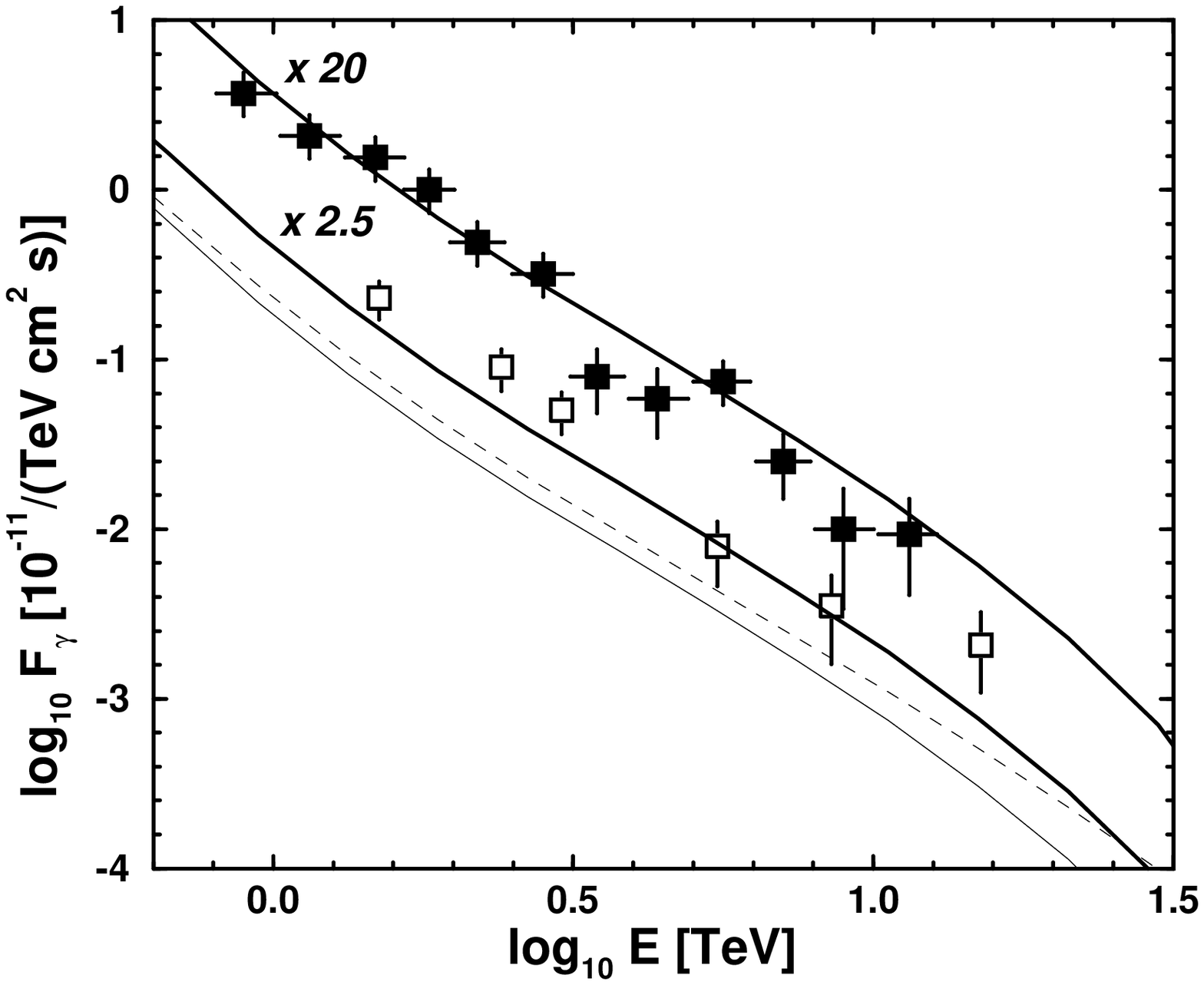,width=12cm,height=10cm}}
\mycaption{{\bf Fig.6:}
Comparison of predicted and observed flux density spectrum
in the multi-TeV range for Mrk~501.  
The thin solid line shows the spectrum published
in \cite{man96}.  The upper thick solid lines
show this spectrum scaled to the 300~GeV flux levels 
during the two observation epochs where the air-Cerenkov
data 
indicated by the
solid and open symbols were obtained with the HEGRA telescopes.
The dashed line shows the spectrum without the effect of the assumed
marginal interagalactic gamma ray attenuation due to interactions
of the gamma rays with metagalactic infrared radiation.
More recent HEGRA observations with higher statistical significance
show some downward curvature in the 10 TeV range which
may be attributed to a stronger
attenuation which is in line with new analyses of COBE data
and ISO
galaxy counts in the Hubble Deep Field \cite{Dwek99}.} 
\end{figure*}

The observed multi-TeV spectrum is 
modified by the quasi-exponential pair attenuation due to collisions
of the gamma rays with photons from the infrared background.  A recent
evaluation of this background  based on direct measurements obtained
from COBE data in the far-infrared, and inferred as a lower limit from number
counts based on ISO observations of the HDF shows that this
effect compensates the shallow downward curvature discovered
by the HEGRA collaboration in the spectrum of Mrk~501 \cite{Konol99}.
If the gamma rays were
due to inverse-Compton scattering, the shallow curvature is 
difficult to understand for a number of reasons given in ref. \cite{mannheim98}.
The most important of them is that one must expect the accelerated
electrons not to be able to reach energies much higher than 10~TeV.
An inverse-Compton spectrum produced by these electrons would
therefore have to show significant curvature approaching this
maximum energy {\em adding} to the inevitable curvature due to
gamma ray interactions with the infrared background photons.
There are some rumours that quantum gravity effects could possibly
suppress pair production over intergalactic distances, but that
remains highly speculative.  I consider the agreement between the proton blazar prediction
and observation very promising for the model,
albeit minor
discrepancies must be expected, since the proton blazar model
is highly simplified in order to avoid too many free parameters and to be predictive.
Therefore I take the freedom to speculate about the emissions associated
with the gamma rays, viz. cosmic rays and high-energy neutrinos in the following
sections.

\section{Neutrino and cosmic ray predictions}

The photo-production of pions leads to the emission of neutrons 
and neutrinos.  The neutrons decay to protons, and such
extragalactic cosmic rays suffer energy losses traversing the
microwave background \cite{RB93}.  At an observed energy of $10^{19}$~eV,
the energy-loss distance is 
$
\lambda_{\rm p}\sim 1~{\rm Gpc}
$ owing to 
pair production.  This distance corresponds to a redshift $z_{\rm p}$
determined by 
$
\lambda_{\rm p}=(c/ H_\circ)
\int_0^{z_{\rm p}} {dz/[(1+z) \bar E(z)]}
$
where 
$
\bar E(z)=\left[\Omega(1+z)^3+\Omega_R(1+z)^2+\Omega_\Lambda\right]^{1\over 
2}
$
with $\Omega+\Omega_R+\Omega_\Lambda=1$. 
Almost independent on cosmology, the resulting value for $z_{\rm p}$
is given by 
$
z_{\rm p}= h_{50}/(6-h_{50})\simeq 0.2 h_{50}
$
where $h_{50}=H_\circ/50~$km~s$^{-1}$~Mpc$^{-1}$.
Therefore, when computing the contribution of extragalactic sources
to the observed cosmic ray flux above $10^{19}$~eV, only sources
with $z\le z_{\rm p}$ must be considered. 
Assuming further that extragalactic
sources of cosmic rays and neutrinos
are homogeneously distributed  with a monochromatic
luminosity density
$\Psi(z)\propto (1+z)^{3+k}$
where $k\sim 3$ for AGN \cite{BT98},
their contribution to the energy density of a
present-day diffuse isotropic background is given by
\begin{equation}
u(0)=
\int_0^{z_{\rm m}} \Psi(z) (1+z)^{-4} {dl\over cdz} dz=
{\Psi(0)\over H_\circ} 
\int_0^{z_{\rm m}} {(1+z)^{\rm k}dz\over {(1+z)^2 \bar
E(z)}}
\end{equation}
where $z_{\rm m}=2$ denotes the redshift of maximum luminosity
density.
The factor
$(1+z)^{-4}$ accounts for the expansion of space and the redshift
of energy. For a simple analytical estimate of the effect
of energy losses on the proton energy density at $10^{19}$~eV,
we collect only protons from sources out to the horizon redshift 
$z_{\rm p}\sim 0.2$
for $10^{19}$~eV protons, whereas neutrinos are collected from sources
out to the redshift of their maximum luminosity density $z_{\rm m}$.
This yields the energy density ratio for neutrinos at an observed
energy of $\sim 5\,10^{17}$~eV and protons at $10^{19}$~eV
\begin{equation}
{u_\nu(0)\over u_{\rm p}(0)}=
{\xi\int_0^{z_{\rm m}} (1+z)^{k-2}/\bar E(z) dz\over
\int_0^{z_{\rm p}} (1+z)^{k-2}/\bar E(z)dz}\sim 2-3
\end{equation}
using $\xi\sim 0.3$ from decay and interaction kinematics,
and considering 
an open Universe with $\bar E(z)=(1+z)$ and a closed
one with $\bar E(z)=(1+z)^{3/2}$.  

Fig.~6 shows exact energy-dependent results for $\Omega=1$ 
from a full Monte-Carlo
simulation employing the matrix doubling method of 
Protheroe \& Johnson \cite{PJ96}
and using the model A neutrino spectrum
from the original work \cite{M95}.  The associated gamma ray flux
corresponds to the observed background flux above 100~MeV\footnote{
A recent paper by Waxman and Bahcall \cite{WB99}
refers to the neutrino flux
from model B in the original work which was given only to demonstrate
that hadronic jets cannot produce a diffuse gamma ray background
with an MeV bump (as measured by Apollo and which is
now known to be absent from a COMPTEL analysis) without over-producing
cosmic rays at highest energies.}.  The neutrino flux is consistent with
the bound given in ref.~\cite{WB99}, although it is possible to
have extragalactic neutrino sources of higher neutrinos fluxes
without violating the observed cosmic ray data as a bound \cite{MRP99}.
Note that there are a few cosmic ray events at energies above $10^{20}$~eV
which are difficult to reconcile with an origin in extragalactic 
radio sources, since the radio galaxies are typically at such large distances that
pion production quenches their spectrum above $10^{19.5}$~eV.
However, cosmic ray particles from the few closest radio galaxies
deflected by magnetic fields could possibly explain these events.
If not, they might originate from the decay of still higher energy particles,
such as the gauge bosons produced at cosmic strings \cite{Sigl94}
indicating new physics.

\begin{figure*}
\centerline{\psfig{figure=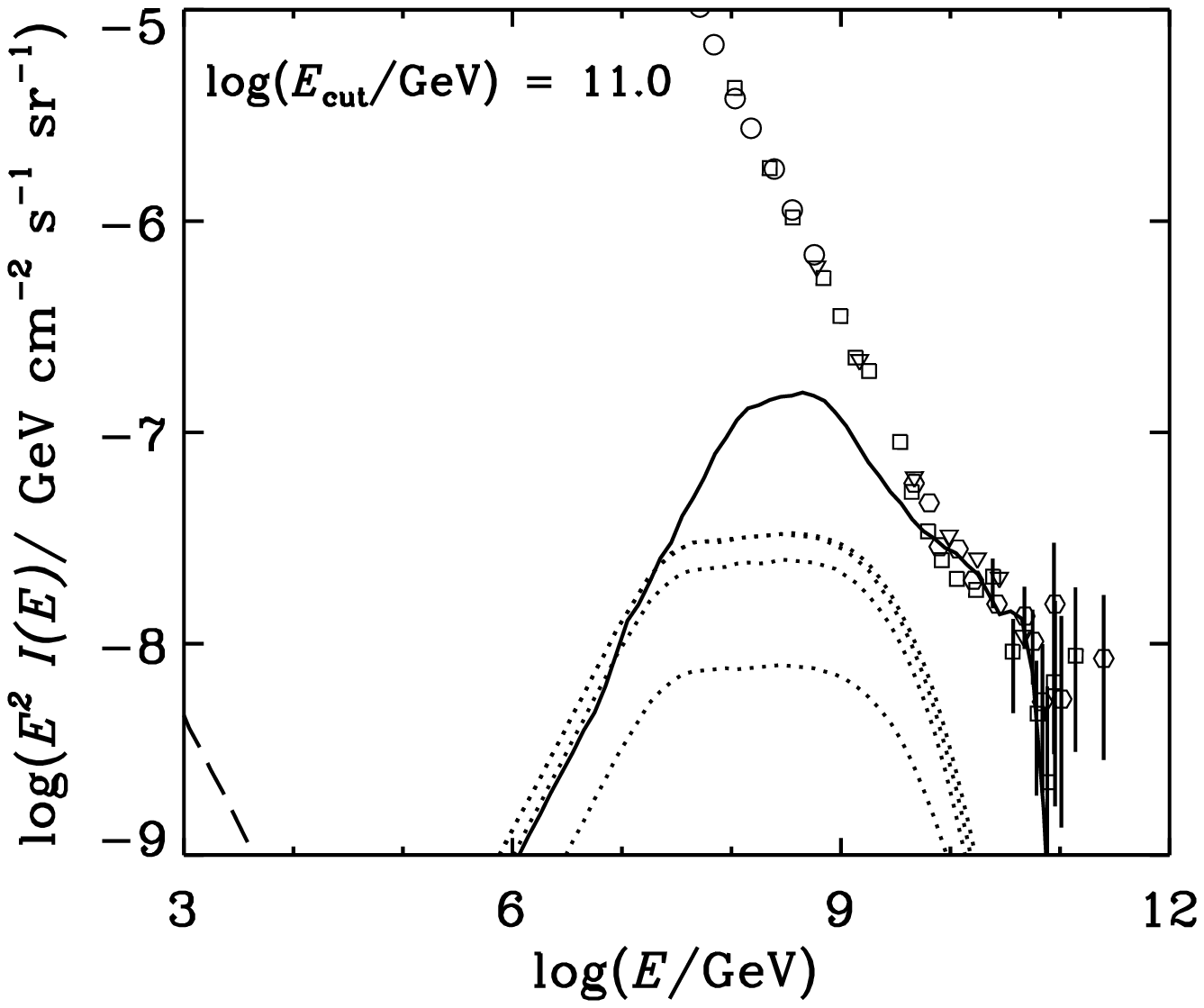,width=10cm,height=8cm}}
\mycaption{{\bf Fig.~7:}
Comparison of proton (solid line) and neutrino fluxes (dotted lines,
from top to bottom $\nu_\mu,~\bar\nu_\mu$, and $\nu_e$) from the proton blazar
model 
(Monte-Carlo computations and figure by R.J. Protheroe). 
The open symbols represent the observed cosmic ray flux.}
\end{figure*}

\section{Neutrino oscillations and event rates}

The neutrino flux shown in Fig.~7 corresponds to
a very low muon event rate even in a km$^2$ detector which
is of the
order of 1~event per year and per steradian.  This event rate
could be increased if there is additional neutrino production
due to pp-interactions of escaping nucleons diffusing through
the host galaxies which is difficult to predict due to their unknown
magnetic fields and turbulence level.
The reason for the low rate is that the neutrino spectrum is extremely
hard, a differential proton spectrum of index $s_{\rm p}=2$ photo-producing
pions in a synchrotron photon target also with differential index 
$s_{\rm syn}=2$
yields a differential neutrino spectrum of index $s_\nu=1$ 
($dN/dE\propto E^{-s}$) up to some very high energy.  The spectrum
may be more shallow, if the target photons have a spectral index of
0.7 as suggested for Mrk~501, thereby increasing the number
of lower energy neutrinos while keeping the bolometric flux the same.
Because of the long lever arm from
$10^{6.5}$~TeV to 1~TeV, a factor of $\sim 100$ increase in
the event rate would result from this effect.

At this point the discovery of neutrino mass announced by the
Super-Kamiokande collaboration \cite{SuperK98} comes in changing the situation
in a major way.  A deficit of atmospheric muon neutrinos was observed
with Super-Kamiokande at large zenith angles with the most likely
explanation being a full-amplitude oscillation of muon flavor eigenstates to
tauon flavor eigenstates across the Earth at GeV energies.
While this would make long-baseline experiments searching for the
appearance of the tauon in a muon neutrino beam with laboratory beams
extremely difficult (if not impossible), it qualifies the expected
extragalactic sources of muon neutrinos as an ideal neutrino beam.
The energies are high enough to
produce tauons on the mass-shell and the distance large enough
to obtain full mixing.  Since tauons decay before interacting in
the Earth and since the Earth is opaque to tau neutrinos above
$\sim$~100~TeV, a fully
mixed extragalactic muon neutrino beam must initiate tauon cascades
in the Earth shifting the tauon neutrino flux down to energies of 
$\sim$~100 TeV \cite{HS98} and obliterating the Earth-shadowing effect 
\cite{G96} that makes the muon solid ange very narrow at high energies. \\

The neutrino oscillations would have another important consequence.
Current data suggested a maximal mixing between muon and tauon
flavors and a mass difference given by 
\begin{equation}
\Delta m_{\mu\tau}^2=m_{\nu_\tau}^2-
m_{\nu_\mu}^2=5\times 10^{-3}~{\rm eV^2}
\end{equation}
The mass difference between electron and muon
flavored neutrinos inferred from the solar neutrino deficit
is orders of magnitude less, and this implies that the neutrino
masses could be highly degenerate if they are at the eV level.
Using the limit $m_{\nu_e}<5$~eV, 
the maximum allowed combined neutrino mass would
be 
\begin{equation}
m_\nu\approx 3 m_{\nu_e}<15~{\rm eV}.
\end{equation}
Inserting this into the Cowsik-McClelland bound one
obtains the maximum contribution to the (hot) dark matter
of the Universe
\begin{equation}
\Omega_\nu< {15\over 91.5}h^{-2}\approx 0.4 \left(h\over 0.65\right)^{-2} .
\end{equation}
Due to free streaming of the neutrinos at the time of recombination,
density fluctuations of this dark matter component would be wiped
out on scales less than
\begin{equation}
\lambda_\nu=30 \left(h/0.65\right)^{-2}~{\rm Mpc}
\end{equation}
and would therefore have no consequence for the dark matter inferred
from studies of galaxy halos or galaxy clusters which yield $\Omega=0.3\pm 0.1$
\cite{Bahcall99}.  
Structure formation simulations exclude $\Omega_\nu$ to
be larger than 0.15 \cite{White83}
which is in agreement with the above upper limit
and both evidences together rule out neutrinos as the
dark matter which could supply a critical mass to the Universe.

\section{Discussion and summary}

The paper highlights in a personally
biased way on a few developments in high-energy astroparticle
physics, 
rather than to give a review of all the activities in the field which
encompass a much wider scope from the origin of cosmic rays to
quantum gravity and which involve a truly impressive number of
experimental efforts.  From my point of view, the identification
of one component of dark matter using the discovery of the atmospheric
neutrino anomaly with Super-Kamiokande represents a major achievement,
and I have highlighted its consequence that neutrinos cannot close
the Universe.  Furthermore, this discovery is not sufficient to
prove the neutrino oscillation hypothesis conclusively.  An experiment
is needed which shows the appearance of the tau lepton in a beam
of muon neutrinos, and I have shown that neutrinos due to proton
acceleration in extragalactic sources would be ideally suited
as a beam for a tau-appearance experiment in one of the major
neutrino telescopes which are under construction.  From an
astrophysical point of view, the oscillations are also
important for another reason, since
they remove the Earth shadowing effect
which suppresses the response of a neutrino telescope to 
extraterrestrial sources of very high-energy neutrinos.  This
makes the discovery of neutrinos from radio-loud
AGN much more probable in a few years.\\

Other kinds of dark matter such as right-handed neutrinos,
supersymmetric
particles, magnetic monopoles, or other topological
defects and their associated gauge bosons, are likely to be very massive.
It is a formidable task to find other than gravitational evidence
for this dark matter due to the electromagnetic (photons), weak (neutrinos),
and strong (protons) couplings of its decay and annihilation products.
Inevitably this task requires to understand all the astrophysical background
radiations at high energies. The astrophysics governing this
energy domain itself has proven to be a fascinating realm.
A good example is the discovery of multi-TeV emission from nearby 
blazars using the HEGRA imaging air-Cerenkov telescopes.  The
observational findings frustrate the worldwide
elite in theoretical astrophysics and seem to provide key insights
into the extraordinary physics driven by weakly accreting black holes.\\

If the recent determinations of the infrared background
from ISO galaxy counts (lower limits) and COBE/DIRBE and FIRAS are
correct, then the steepening in the multi-TeV spectrum of
Mrk~501 observed with the HEGRA air-Cerenkov telescopes
 is due to the predicted gamma ray attenuation in collisions with these
infrared background photons.  Predictions of the attenuation process 
for sources at higher redshifts depend strongly on the
evolution of the metagalactic optical-UV radiation field. 
It will therefore soon be possible to independently probe the 
star formation history using 10-100~GeV
gamma rays from extragalactic high-redshift 
sources when lower threshold gamma ray telescopes such as
MAGIC and GLAST are available.  
The method is sensitive to truly diffuse photons between the
galaxies and, when compared with calculations based on the
observed star formation rate in early galaxies,
allows to test the hypothesis whether the background photons originate from
the stars or other sources (possibly connected with the
reionization epoch).  \\

The multi-TeV spectra from nearby blazars prediced on the basis of the proton
blazar model are in accord with the observations if the effect of pair
attenuation due to the extragalactic infrared background is taken into account.
This is surprinsing, since the model gives stationary spectra, whereas
the observed flux is highly variable.  Nevertheless, the spectra seem
to remain rather constant in the multi-TeV domain.  There are several
properties which are not explained in the framework
of a stationary quasi-homogeneous 
model by contruction, such
as the apparently different variability pattern of Mrk~421 in the GeV
and TeV ranges.   This does not argue against the proton acceleration
hypothesis, since one could  construct inhomogeneous,
non-stationary models to explain this phenomenology.  The same
is true for the problem of short-term variability, which may
require to describe the passage of shocks traveling through
narrowly-spaced inhomogeneities.
It is amazing, however, that a general feature
expected from models based on electron acceleration is certainly not
seen in the data of Mrk~501, viz. the change of the upper cutoff energy
with varying flux.  This can only mean that the cutoff is indeed due
to intergalactic attenuation and that the electron maximum energy
is much higher than 20~TeV which is difficult to understand in the
presence of synchrotron and inverse-Compton losses.  Another peculiar
finding is that the multi-TeV emission in Mrk~501 is accompanied
by 100~keV synchrotron photons in some epoches, but not in all of them.  
Moreover,
in Mrk~421 100~keV synchrotron flares have not been observed in spite
of flaring TeV emission. So the story is not simple and likely to 
continue, brainstorming is always desired when confronted with a new
phenomenon.\\

An interesting corollary from hadronic models of extragalactic gamma ray
sources is that they would also emit cosmic rays and neutrinos at about
equal luminosities (within a factor of a few).
If the cosmic ray flux emitted by hadronic
accelerators equals that of the observed cosmic rays above $10^{19}$~eV, the
associated gamma ray power from these sources is enough to produce the observed
extragalactic gamma ray background above about 100 MeV.  The gamma ray power is
larger than that in cosmic rays, since the cosmic rays lose energy traversing
the low-energy background radiation fields and most sources have high redshifts.
It has been proposed in ref. \cite{WB97} that GRBs are responsible for
the highest energy cosmic rays, but more recent GRB observations
indicate that most of them have high redshifts which is expected if they
trace star formation.
The strong evolution of the GRB luminosity density would 
then rule out GRBs as possible
sources of the highest energy cosmic rays,
since their cumulative gamma ray flux
is far below the putatively extragalactic gamma ray flux.  
Although the muon event
rate in neutrino telescopes from hadronic extragalactic gamma ray
sources supplying the highest energy cosmic rays
is low, neutrino oscillations lead to tau cascades
canceling the Earth shadowing effect thereby increasing the detection
probability.   Tau lepton appearance in the neutrino
telescopes would constitute the first direct measurement of the tauon
(which so far has only been inferred from momentum conservation) 
and would prove the neutrino oscillation hypothesis.\\


\begin{thebibliography}{999}

\bibitem{Turner99}
M.S. Turner, in: {\em Physica Scripta}, Proceedings of the Nobel
Symposium, Particle Physics and the Universe, Enkoping, Sweden,
August 20-25 (1998).

\bibitem{Gaisser95}
T.K. Gaisser, F. Halzen, T. Stanev, {\em Phys. Rep.} {\bf 258} (1995) 173.

\bibitem{Perl98}
S. Perlmutter, et al., {\em Astrophys. J.} (1999) in press.

\bibitem{Schmidt98}
B. Schmidt, et al., {\em Astrophys. J.} {\bf 507} (1998) 46.

\bibitem{Sigl94}
G. Sigl, D.N. Schramm, P. Bhattacharjee, {\em Astropart. Phys.}
{\bf 2} (1994) 401; P. Bhattacharjee, G. Sigl, {\em Phys. Rep.}
(1998) submitted.

\bibitem{MAGIC}
The MAGIC Telescope Project:  http://hegra1.mppmu.mpg.de:8000/.

\bibitem{GLAST}
The GLAST Project: http://glast.gsfc.nasa.gov/LHEA/.

\bibitem{Aharon97}
F. Aharonian, et al. (HEGRA Collaboration), {\em Astron. Astrophys.}
{\bf 327} (1997) L5.

\bibitem{Konol99}
A. Konopelko, et al. (HEGRA Collaboration), in: {\em VERITAS Workshop
on TeV Astrophysics of Extragalactic Sources}, eds. M. Catanese,
J. Quinn, and T. Weekes, to be published in Astropart. Phys. (1999).

\bibitem{Dwek99}
O.C. de Jager, E. Dwek, {\em Astropart. Phys.} (1999) submitted.

\bibitem{SuperK98}
Y. Fukuda, et al.
(Super-Kamiokande Collaboration), {\em Phys. Rev. Lett.} {\bf 81} (1998) 1562.

\bibitem{Rees98}
M.J. Rees, J. Silk,  {\em Astron. \& Astrophys.} (1998) {\bf 331}, L1.

\bibitem{Sanders89}
D.B. Sanders, et al., {\em Astrophys. J.} {\bf 347} (1989) 29.

\bibitem{Gruber92}
D.E. Gruber, {\em The X-ray Background}, eds. X.~Barcons and
A.C.~Fabian, Cambridge U.P. (1992) p. 44.

\bibitem{RS91}
S. Rawlings, S. Saunders, {\em Nature} {\bf 349} (1991) 138.

\bibitem{MP96}
P. Madau, E.S. Phinney, {\em Astrophys. J.} {\bf 456} (1996) 124.

\bibitem{Salamon98}
M.H. Salamon, F.W. Stecker, {\em Astrophys. J.} {\em 493} (1998) 547.

\bibitem{gould66}
R.J. Gould, G. Shr\'eder, {\em Phys. Rev. Lett.} {\bf 16} (1966) 252.

\bibitem{jauch76}
J.M. Jauch, F. Rohrlich, {\em  The Theory of Photons and Electrons}, 
Springer-Verlag, Berlin (1976).

\bibitem{fixsen98}
D.J. Fixsen, et al., {\em Astrophys. J.} {\bf 490} (1997) 482.

\bibitem{franceschini98}
A. Franceschini, et al., in: {ESA FIRST Symposium}, ESA SP 401
(1997) astro-ph/9707080.

\bibitem{mannheim98}
K. Mannheim, {\em Science} {\bf 279} (1998) 684.

\bibitem{biller96}
S.D. Biller, {\em Astropart. Phys.} {\bf 3} (1996) 385.

\bibitem{Madau}
P. Madau, in: {\em 7th Annual October Astrophysics Conference in Maryland,
``Star Formation Near and Far''} (1996) astro-ph/9612157.

\bibitem{Chiang}
J. Chiang, R. Mukherjee, {\em Astrophys. J.} {\rm 496} (1998) 752.

\bibitem{Dar99}
A. Dar, A. De R\'ujula, N. Antoniou, {\em Phys. Rev. Lett.} (1999) submitted.

\bibitem{MKB91}
K. Mannheim, W. Kr\"ulls, P.L. Biermann, {\em Astron. \& Astrophys.}
{\bf 222}(1991) 222; K. Mannheim, P.L. Biermann, {\em Astro. \& Astrophys.}
{\bf 333}(1992) L21; K. Mannheim, {\em Astron. \& Astrophys.}
{\bf 269} (1993) 67.

\bibitem{man96}
K. Mannheim, S. Westerhoff, H. Meyer, H.H. Fink,
{\em Astron. Astrophys.} {\bf 315} (1996) 77.

\bibitem{RB93}
J.P. {Rachen} and P.L. {Biermann}, {\em Astron. \& Astrophys.} {\bf 272} 
(1993)  161.


\bibitem{BT98}
B.J. {Boyle} and R.J. {Terlevich}, {\em Mon. Not. Roy. Astro. Soc.}
{\bf 293} (1998) L49.

\bibitem{PJ96}
R.~J. {Protheroe} and P.~A. {Johnson}, {\em Astropart. Phys.} {\bf 4} (1996)
253.

\bibitem{WB99}
E. Waxman,  J. Bahcall, {\em Phys. Rev. D} (1999) accepted.

\bibitem{M95}
K. {Mannheim}, {\em Astropart. Phys.} {\bf 3} (1995)  295.

\bibitem{MRP99}
K. Mannheim, J.P. Rachen, R.J. Protheroe, {\em Phys. Rev. D.}
(1999) submitted.

\bibitem{HS98}
F. Halzen, D. Saltzberg, {\em Phys. Rev. D.} {\bf 81} (1998) 4305.

\bibitem{G96}
R. {Gandhi}, C. {Quigg}, M.~H. {Reno}, and I. {Sarcevic}, {\em
Astropart. Phys.} {\bf 5} (1996)  81.
  
\bibitem{Bahcall99}
N.A. Bahcall, 
in: {\em Physica Scripta}, Proceedings of the Nobel
Symposium, Particle Physics and the Universe, Enkoping, Sweden,
August 20-25 (1998).

\bibitem{White83}
S.D.M. White, C. Frenk, M. Davis, {\em Astrophys. J.} {\bf 274} (1983) L1.

\bibitem{WB97}
E. Waxman, J. Bahcall, {\em Phys. Rev. Lett.} {\bf 78} (1997) 2292.


\end{thebibliography}
\end{document}